# Configured Grant for Ultra-Reliable and Low-Latency Communications: Standardization and Beyond

Majid Gerami and Bikramjit Singh

Ericsson Research

*Abstract: Uplink configured Grant allocation has been introduced in 3rd Generation Partnership Project New Radio Release 15. This is beneficial in supporting Ultra-Reliable and Low Latency Communication for industrial communication, a key Fifth Generation mobile communication usage scenario. This scheduling mechanism enables a user with periodic traffic to transmits its data readily and bypasses the control signaling entailed to scheduling requests and scheduling grants and provides low latency access. To facilitate ultra-reliable communication, the scheduling mechanism can allow users to transmit consecutive redundant transmissions in a pre-defined period. However, if the traffic is semi-deterministic, the current standardized configured grant allocation is not equipped to emulate the traffic as the configured grant's period is pre-configured and fixed. This article describes the recent advancements in the standardization process in Release 15 and 16 for configured grant allocation and the prospective solutions to accommodate semi-deterministic traffic behavior for configured grant allocations.*

*Index Terms— Fifth Generation (5G), New Radio (NR), Ultra-Reliable Low Latency Communications (URLLC), Configured Grant (CG)*

## 1  INTRODUCTION

The Fifth Generation (5G) of mobile communication was launched around the world in the early 2020s and now discussion on standardization and research are ongoing in 5G advanced and 6G. One of the key usage scenarios that are targeted is Ultra-reliable and low latency communications (URLLC) for beyond International Mobile Telecommunications-Advanced services [1]. This enables real-time control and automation of dynamic processes for vertical applications such as factory automation, transport industry (e.g., remote driving), smart grids, and electrical power distribution. These services may require the reliability of an order of 99.999 % and low radio latency down to 1 ms. To support such services with extreme requirements, the 5G New Radio (NR) networks have been architected in Third Generation Partnership Project (3GPP).

To fulfill the stringent requirements on low-latency access for URLLC, especially for uplink transmission, it is important to eliminate the delays due to Scheduling Request (SR) from the User Equipment (UE) to the Next-generation Node B (gNB) and the subsequent grant from the gNB to the UE [2]. Further, it also improves the uplink transmission's reliability by discounting the probability failure of SR transmitted over an uplink control channel, i.e., Physical Uplink Control Channel (PUCCH), and grant over a downlink control channel, i.e., Physical Downlink Control Channel (PDCCH). The data is transmitted readily on the allocated resource. Hence, instead of dynamically scheduling a transmission, the radio resources are allocated periodically approximately mirroring the UE's periodic traffic needs. This uplink scheduling feature without requiring SR and grant is denoted as Configured Grant (CG) in NR releases [3-4]. Further, coupled with new numerology adopted in NR [5], e.g., with Subcarrier Spacing (SCS) 120 kHz in Frequency Range 2 (FR2), the slot length is 125 μs and the possibility to transmit on fewer symbols in a slot, CG can provide ultra-low latency access.

Likewise, NR CG, Semi-Persistent Scheduling (SPS) based periodic allocation in both uplink, *SPS-ConfigUL* and downlink, *SPS-ConfigDL* [6] resources is already feasible in Fourth Generation (4G)

Long Term Evolution (LTE), e.g., providing Voice over Internet Protocol (VOIP) services. However, 5G URLLC services posed much stricter latency and reliability requirements and 4G inspired uplink SPS allocation cannot be used as such. The primary reason is that 4G operates with 15 kHz SCS only, whereas higher numerology is available in 5G, which furnishes a much smaller Orthogonal Frequency-Division Multiplexing (OFDM) symbol size in the time domain.

In NR Release 15 and 16 [3-4], various aspects of CG are covered focusing on small packet transmission, e.g., < 50 bytes [1] to provide guaranteed reliability under a strict latency budget. The underlying assumption is deterministic arrivals. However, if the scenario is expanded to semi-deterministic arrivals or high bit-rate application, e.g., Extended Reality (XR) [7] or spectrum efficient networks, then possible work remains in further stages of the NR CG evolution.

This article provides an overview of the status of NR CG in 3GPP and the possible enhancements for its evolution. In the next section, we elaborate on the 3GPP standardization aspects of CG and describe some notable Physical (PHY) and Medium Access Control (MAC) layer features for CG resource allocation. Next, we outline some prospective solutions to cater for non-deterministic arrivals, where current CG is equipped to deal with moderate traffic variations.

## 2 CONFIGURED GRANT STANDARDIZATION

3GPP has adopted two tracks for CG specifications and enhancements in Release 15 and 16, namely, CG in (a) NR Release 15 and 16 [3-4, 8] and (b) NR-Unlicensed (NR-U) Release 16 [4, 8, 9]. For the CG operation in NR licensed spectrum, the enhancements aim for the extreme requirements of URLLC. On the other hand, NR-U focuses on reducing the effect of Listen-Before-Talk (LBT) failure for a CG operation in the unlicensed or shared spectrum. The working phase in Release 15 and 16 has already finished and now in Release 17, work is ongoing to harmonize these two CG variants [10].

In CG, the time-frequency resources are assigned periodically where the UE transmits on these pre-assigned resources without the need for an SR. These resources can be configured with two types of procedural signaling, which are Type 1 and Type 2 CG. For Type 1 CG, the resource allocation is provided by higher layer Radio Resource Control (RRC) parameters. For Type 2 CG, the resource allocation is provided jointly by RRC parameters and by Downlink Control Information (DCI). Type 1 CG does not require DCI, hence, the precious PDCCH resources are saved, unlike in Type 2. On the other hand, resource allocation and adjustments to the allocations using DCI signaling are much faster in comparison to RRC signaling, which makes Type 2 more effective for URLLC especially during the activation, or changing of CG allocations. In the subsequent sections, the properties of CG Type 1 and Type 2 are considered for the discussion focusing on PHY and MAC layers aspects [3-4, 8].

### 2.1 Transmission resource

For Type 1 CG, the resource assignment is done during configuring the RRC parameters. The parameters *timeDomainOffset*, *timeDomainAllocation* and *frequencyDomainAllocation* parameters in *ConfiguredGrantConfig* information element provide the information to a UE about the time and frequency domains of resource allocation. For Type 2 CG, the resource assignment is done via DCI which is coupled with RRC parameters. In DCI, the time domain allocation is derived from the value in the DCI's Time Domain Resource Assignment (TDRA) field index to a row in an allocation table in RRC. The indexed row defines the slot offset, the start and length indicator value

SLIV (indicating start symbol $S$ and allocation length $L$), the Physical Uplink Shared Channel (PUSCH) mapping type and the number of repetitions to be applied in the PUSCH transmission.

The Frequency Domain Resource Assignment (FDRA) field in the activation DCI represents the frequency domain allocations. For both Type 1 CG and Type 2 CG, there are three types of frequency resource allocations, namely Type 0, Type 1, and Type 2. In FDRA Type 0, the frequency allocation is specified by the Resource Block Group (RBG) bit-map. Each RBG contains several contiguous Resource Blocks (RBs). In FDRA Type 1, the frequency allocation is over contiguous virtual RBs, the start and length of RB allocations are indicated in the activation DCI. In FDRA Type 2, which is primarily used for NR-U, the frequency allocation is indicated through interlaced RBs.

## 2.2 *Transmission repetition*

To improve CG transmission reliability, the allocation can be supported with $K$ consecutive redundant allocations or Transmission Occasions (TOs) within a pre-defined period allowing multiple Transport Blocks (TBs) to be transmitted repeatedly in the period. The length of the period can be established as per the latency budget of the transmission, where the TOs are assigned with a definitive Redundancy Version (RV) pattern. There are four types of RVs defined in NR, RV0, RV1, RV2, and RV3. Data encoded by RV0 and RV3 are self-decodable. Currently, in NR, three RV patterns are allowed, {0, 0, 0, 0}, {0, 3, 0, 3} and {0, 2, 3, 1}. In Release 15, transmission is allowed to begin only at first TO. However, Release 16 has introduced a Boolean type higher layer parameter *startingFromRV0-r16* that configures transmission to begin at the first TO if the parameter is set off, however, if turned on, the transmission may begin at any RV0 TOs in the pattern except for $K \geq 8$ repetitions. The flexible beginning of transmission provides flexibility in the trade-off between reliability and latency. By setting '*off*' the parameter *startingFromRV0-r16* the transmission the reliability is guaranteed as the transmission is always with $K$ repetitions or no transmission at all. This however increases the latency, since UE would need to wait for the next period to begin its transmission if data arrives in the middle of K repetitions. On contrary, setting '*on*' the parameter *startingFromRV0-r16* the latency between the data arrival and the beginning of the corresponding data transmission is reduced as it can be initiated with the nearest TO assigned with RV0 in the same period.

In NR Release 15, the repetitions over consecutive slots are termed as slot aggregation, e.g., Fig. 1-a depicts a CG transmission with slot aggregation $K = 2$ and CG periodicity $p = 4$ slots. When slot aggregation applies to mini-slots, there would be a delay between the repetitions as a UE occupies the same mini-slot in every slot for its consecutive repetitions, see Fig. 1-b. To mitigate this delay, mini-slot repetition is endorsed in NR Release 16, where a short transmission can be repeated within a slot, see Fig. 1-c. This is known as PUSCH Type B repetition in Release 16. Another property of Type B repetition is that the transmission can cross the slot border by splitting into two segmented repetitions before and after the slot border. The same procedure can be employed when a transmission confronts invalid OFDM symbols (OS), e.g., in Time-division duplexing (TDD) pattern. The transmission is split by transmitting over valid symbols and skipped over invalid symbols.

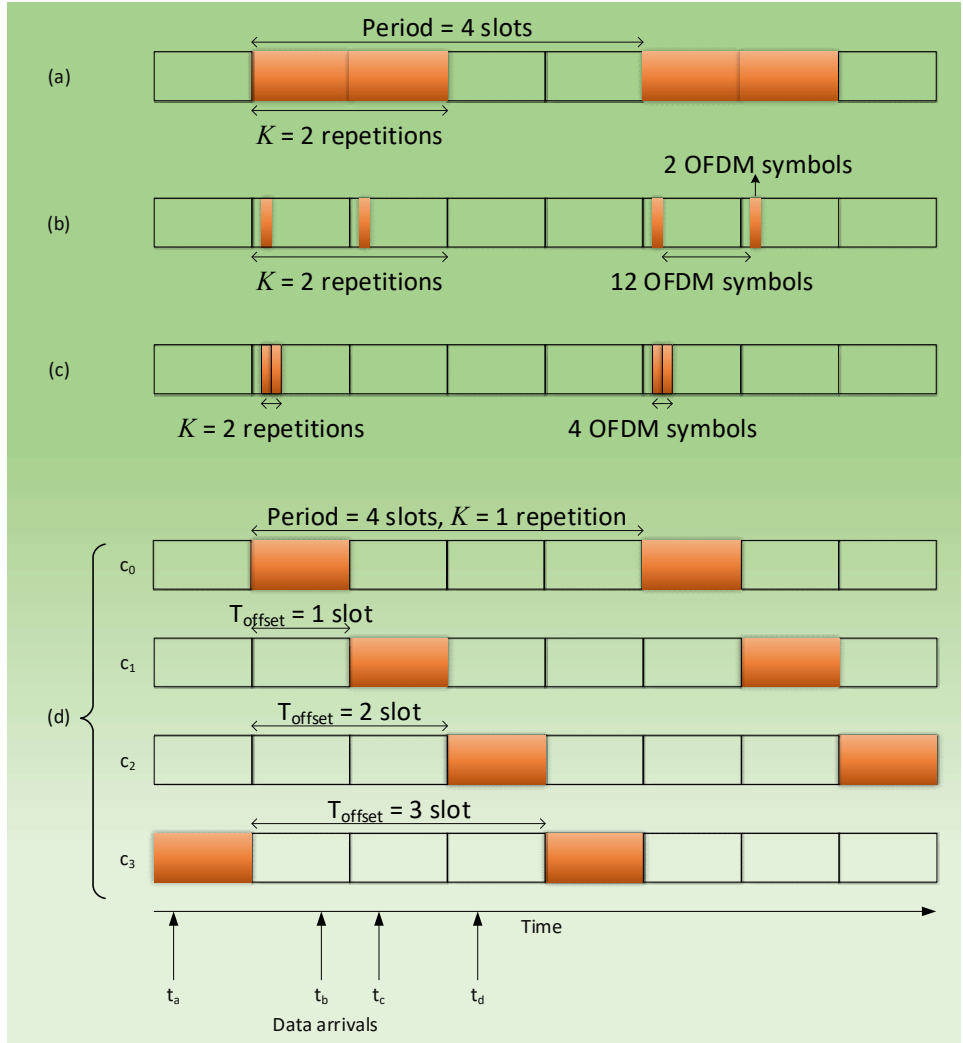

**Figure 1:** Repetition in NR Release 15 and 16. **(a)** Release 15 slot aggregation **(b)** Release 15 repetition of mini-slots **(c)** Additional Release 16 Type B repetition **(d)** Configuring multiple CGs

Further, with CG allocation using Type B repetitions, there are scenarios that the Type B repetitions may result in lower gains in reliability. It means, if there is no need for segmentation, e.g., transmissions over the invalid period or crossing the slot boundary, then the transmission with unnecessary multiple repetition allocations is less reliable than a single repetition over the same resource, which is depicted in Fig. 2. Yet, we note that multiple repetitions allocation has better performance from the latency point of view since more TOs allow UE to have a flexible starting time, especially when more than one user is sharing the same resource.

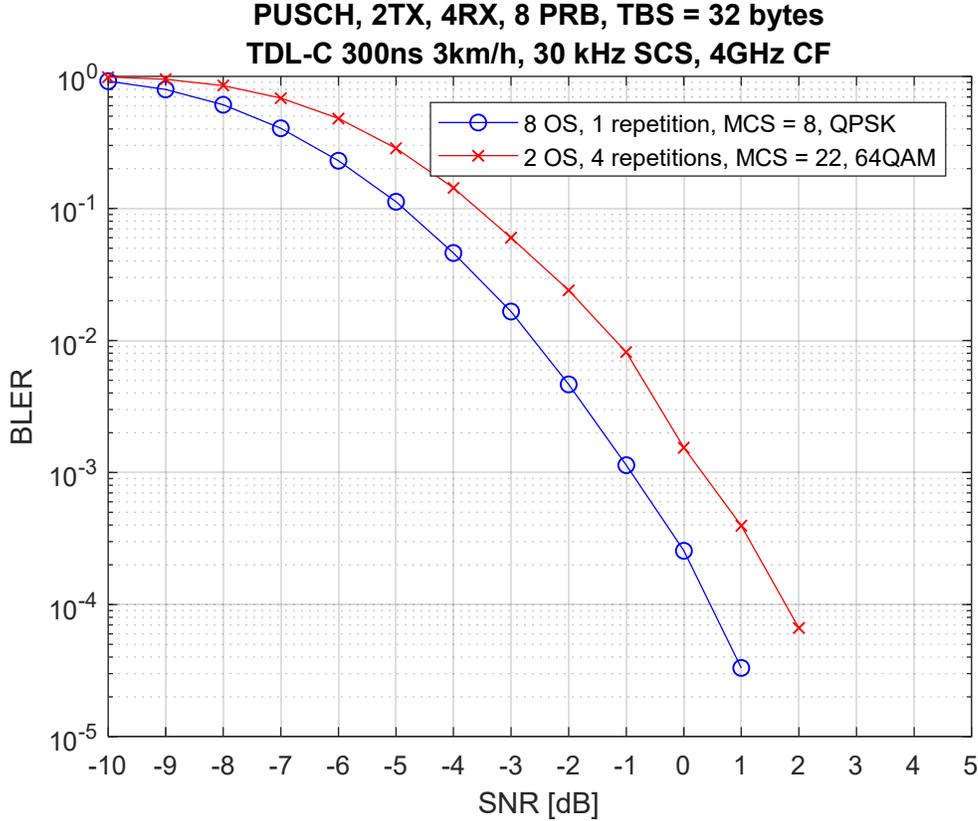

**Figure 2**: BLER performance for a single transmission (of length **8** OS) compared to mini-slot repetition (**4** repetitions of **2** OS).

### 2.3 Number of configurations

In NR Release 15, a UE can be configured with only a single configuration. By single configuration, we mean, a single pattern of periodic uplink allocations is granted to the UE. However, the UE, especially in an industrial environment, can be expected to handle different services, which may have different traffic patterns, packet arrivals, payload size, reliability and latency requirements, etc. To enable several patterns, in Release 16, the UE can be configured with multiple configurations to serve a plurality of traffic with different patterns. Another inherent advantage of configuring the UE with multiple configurations is in reducing alignment delay. This is illustrated in Fig. 1, where a UE is allocated with 4 CGs indicated with CG Identity (ID) $c_0$, $c_1$, $c_2$ and $c_3$. Each CG is configured with the same periodicity $p = 4$ slots and $K = 1$ repetition but different starting TO offset. When the data arrive in the buffer, the UE selects the nearest CG for its transmission, e.g., if data arrives at a time $t_a$, CG $c_0$ is selected, if data arrives at a time $t_b$, CG $c_1$ is selected, and so on. On the contrary, having a single CG only, the UE needs to wait to transmit in the next period whenever data arrives in the middle of a period and there is no TO with RV0 until the end of the period. This can cause increased latency for URLLC services. It must be noted that for a given Hybrid Automatic Repeat Request (HARQ) process transmission, the UE utilizes only one CG and cannot distribute its transmission in multiple CGs.

## 2.4 Activation and release of configurations

For Type 1 CG a configuration is activated by a time offset parameter among RRC parameters. However, the parameters for Type 2 CG are partly configured using RRC and the CG is activated, once a UE receives an activation message in the form of DCI containing resource allocation information. An activation DCI is considered a valid activation message if it is scrambled by Configured Scheduling Radio Network Temporary Identifier (CS-RNTI), and the fields New Data Indicator (NDI), HARQ process ID, RV, and Downlink Feedback Information (DFI) are set to zero. When a UE is configured with multiple CGs, the HARQ process ID field indicates the CG ID(s) and thus is not used for validating activation DCI. Each CG needs thus to be activated separately. Correspondingly, for releasing a CG or a group of CGs, for Type 1 CG, there is an RRC reconfiguration message. For Type 2 CG, there is a release DCI. In Release 16 a group of CGs can be released by a single DCI, while, as already mentioned, activating a group of CGs by a single DCI is not yet adopted in 3GPP standards. UE confirms the (de)activation message(s) through a MAC Control Element (CE).

## 2.5 Retransmission

The initial transmission on CG resource even with repetitions may have a viable failure probability. Hence, if gNB requires the UE to retransmit, it can schedule the retransmission with a dynamic grant by sending a scheduling DCI scrambled by CS-RNTI and NDI bit set to 1. The DCI can be of any DCI format 0-0, 0-1, or 0-2.

## 2.6 PHY priority

Release 16 introduces a two-level priority on a PHY layer, i.e., a field is established in the DCI which labels the grant as low or high priority. However, the priority of CG is indicated as a higher layer parameter, e.g., with RRC, and the priority field in the DCI remains unutilized. The priority plays an important role in selecting or prioritizing the transmissions if a UE has multiple overlapping grants with different priorities, e.g., between two overlapping CGs, the UE cancels the lower priority CG. In Release 16, there is no support for prioritization between overlapping CG and dynamic grants if their PHY priorities are different, and if the priorities are the same, the prioritization will consider numerous other factors including the 16-level Logical Channel (LCH) priority. In the end, MAC generates one TB for the overlapping grants which can be dynamic grants or CG. Release 15 has no PHY layer priority, but rather governs priority choice with an implicit rule where a dynamic grant is always prioritized over a CG, even if the dynamic grant has no data in buffer for transmission.

## 2.7 Unlicensed operation

The unlicensed 5, and 6 GHz bands could potentially offer around 1800 MHz of the spectrum [11]. In controlled environments where the interferences are minimal, an unlicensed spectrum could be potentially utilized to provide URLLC traffic. Thus, 3GPP has stepped in to reshape NR for unlicensed operation, which is defined under NR-U standards. NR-U enables the use of NR features while working under regulations on channel sensing norms such as LBT. For example, in Release 16, NR-U utilizes two modes – Load-Based Equipment (LBE) and Frame-Based Equipment (FBE) for assuring fairness among competing nodes in the unlicensed spectrum. LBE mode is similar to that of Wi-Fi, where a UE transmits if the channel is free and otherwise performs a random back-off. In FBE mode, the resources are divided into Fixed Frame periods (FFPs) and the gNB performs

LBT before each FFP. gNB grabs the FFP if it is not occupied, otherwise, it waits for the next FFP to repeat the procedure. FBE is highly useful in a controlled environment, where the competing nodes are synchronized, or there is less competition among them.

The peculiar case with CG in NR-U is that it has adopted its own track of standardization and diverged from NR Release 16 CG. NR-U CG is configured with the higher layer parameter *cgRetransmissionTimer* which is coupled with many other distinct features such as CG-UCI, autonomous retransmission, UE selects its HARQ ID, RV pattern, etc. The NR-U CG operates with contrasting feedback behavior unlike in NR. If the UE has not received CG-DFI before *cgRetransmissionTimer* expires, the UE assumes Negative ACK (NACK) and can retransmit autonomously in the next TO on the CG. There is also another timer *configuredGrantTimer* (CGT), which limits the maximum number of autonomous retransmission attempts, see Fig. 3. Since the retransmissions are autonomous, a UCI named CG-UCI is included in every CG PUSCH transmission indicating HARQ process ID, RV, NDI, Channel Occupancy Time (COT) sharing information, and Cyclic Redundancy Check (CRC). In NR, UE assumes Acknowledgement (ACK) after the CG timer expires where NACK in the form of a dynamic grant is sent explicitly before the timer when needed. NR-U utilizes CG-DFI, a newly introduced DCI format that carries HARQ-ACK bit-map for all uplink HARQ processes.

The NR-U CG period allocates multiple consecutive TOs which is indicated by higher-layer parameters *cg-nrofPUSCH-InSlot-r16* and *cg-nrofSlots-r16*. The number of TOs can be greater than *K* repetitions and thus multiple HARQ processes are allowed in the CG period, each with *K* repetitions, except the last process, which can have less than *K* repetitions. This is one of the reasons why the HARQ ID selection is not be coupled to the CG in NR-U but rather left to the UE to indicate in CG-UCI.

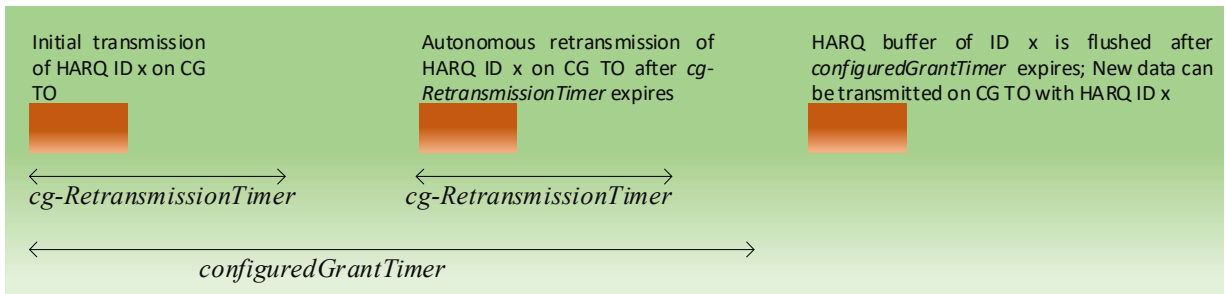

**Figure 3:** NR-U CG autonomous retransmission behavior.

Table I

CG features standardized in Release 15 and 16 for NR and NR-U operation.

| CG features | NR CG | | NR-U CG |
|---|---|---|---|
| | **Release 15** | **Release 16** | **Release 16** |
| **Maximum number of configurations** | One | 12 | 12 |
| **Activation/Release DCI** | DCI format 0-0, 0-1 | DCI format 0-0, 0-1,0-2 | DCI format 0-0, 0-1 |
| **Group Release** | Not Supported | Supported | Not supported |
| **Repetition** | Type A repetition | Type A and B repetition; Type B repetition can cross the slot boundary | Type A and B repetition; Type B repetition cannot cross the slot boundary |
| **PHY priority** | Not Supported | High and low priority | Not Supported; Retransmissions have implicit high priority |
| **ACK feedback** | ACK is implicit | ACK is implicit | ACK is explicit |
| **Autonomous transmission** | Not Supported | Not Supported | Supported |
| **CG-UCI** | Not Supported | Not Supported | Supported |
| **DFI** | Not Supported | Not Supported | Supported |
| **Transmission beginning in a period** | At first TO | At first or any TO with assigned RV 0 | At any TO |
| **HARQ ID determination** | Based on gNB's defined rule | Based on gNB's defined rule | UE chooses and indicates in CG-UCI |
| **RV pattern determination** | Indicated by gNB | Indicated by gNB | UE selects and indicates in CG-UCI |
| **Autonomous retransmission on CG** | Not supported | Not supported | Supported |
| **Number of HARQ processes per CG period** | One | One | More than one is allowed |

## 3 CONFIGURED GRANT ENHANCEMENTS

In this section, we propose some notable CG enhancements to deliver robust URLLC services. New use cases like XR, semi-deterministic services are expected to be a part of the 5G evolution. Currently, Release 17 is already working on an XR study item and proposals related to variable traffic patterns for both uplink video and pose control traffics are being discussed [7]. Thus, we expect that these enhancements target Release 18 or future releases that will deal with scenarios involving capacity traffic with variance, spectrum efficiency, additional spectrum use in the form of NR-U, etc. [15]. In Table II, we list and map the proposed enhancements catering to the potential scenarios. In the later section, these enhancements are discussed in detail.

Table II

Enhanced CG features.

| CG feature | Requirements and scenarios | Comment |
|---|---|---|
| Time-gap between repetitions | Traffic variance | Some repetitions are saved by allocating time-gaps between repetitions resource if the first repetition arrives late |
| Flexible repetition transmission | Traffic variance | The first transmission is done flexibly, and thus all repetitions can be done without skipping |
| Common NACK feedback | Interference-prone environment | If a transmission is misdetected, then a broadcasting NACK would alert such UE for its failed transmission |
| Assistive shared resource | Traffic variance, spectrum efficiency, load improvement | Additional spectrum, e.g., in the form of NR-U can be utilized to transmit failed repetitions |

| Complement TDD pattern | Restrictive TDD pattern, spectrum efficiency, load improvement | An additional carrier can be configured which is in compliment to the TDD pattern of another carrier, and thus all repetitions can be transmitted fully |
|---|---|---|

### 3.1 Time-gap between repetitions

The TOs within a CG period are allocated consecutively at the beginning of a period. If data, also, arrives at the beginning of the period, the UE can utilize all the TOs as needed. However, in practice, data arrival is not fully deterministic. If data arrives later in the period, the UE may miss some or all TOs occurring in the CG period. To diminish the loss of TOs, a suitable time-gap between the TOs can be configured. With the introduction of time-gaps, UE would have a higher probability to transmit on the number of TOs than it would do with no time-gaps, see Fig. 4. For example, if a CG period consists of $N$ slots, and assuming a worst-case scenario of random traffic with data arrival probability in a slot is $1/N$, then the probability of transmission of at least one repetition is $K/N$ where a CG period is allocated with $K$ consecutive TOs for $K$ repetitions. However, if the CG period is allocated with a time-gap of $T$ slots between every TOs, then the probability of transmission of at least one repetition would become $(T(K-1) + K)/N$, which is larger than $K/N$.

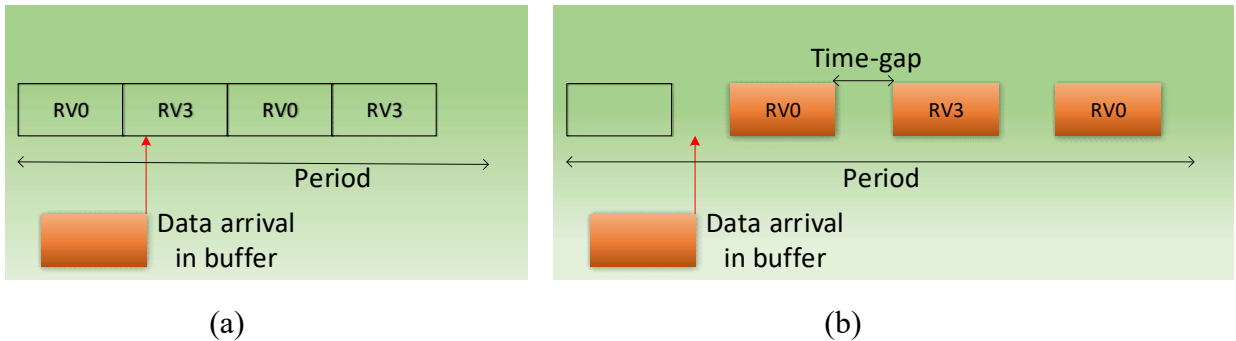

(a)          (b)

**Figure 4:** CG TOs allocation with Release 16 design and enhancements (a) flexible repetition transmissions and (b) time-gap.

### 3.2 Flexible repetition transmission

In NR CG allocation, the TOs with assigned RV0 allocation are fixed. In case of late arrival, this can lead to no or fewer repetition transmissions given the delay in UE arrival and the number of configured RV0 occasions in the period onto which UE can begin its transmission. This can have a substantial adverse impact on reliability, and to subside this impact of fluctuating arrival, one solution is that, instead of tying the RV pattern to the allocated TOs, the RV pattern is associated with the arrival. This means, if a UE begins late in the period, it transmits readily in the nearest remaining occasions based on the RV pattern, see Fig. 4. For example, if a UE begins with transmitting on $i$-th occasion from $K$ allocated occasions, it transmits with RV0 repetition on $i$-th occasion, and the repetitions on the remaining occasions follow the RV pattern. However, as per Release 15 and 16 CG, a UE can only begin with $i$-th occasion if $i$-th occasion is configured with RV0, otherwise, the UE looks for the nearest $j$-th occasion which is configured with RV0 to begin with its transmission repetitions where $i < j < K$. A small analytical comparison can be made to demonstrate the utilization of the allocated occasions in both cases. Let us assume, a UE arrives at $b$-th occasion in a period with $K$ allocated occasions. According to Release 15 and 16 CG, if the RV0 occasion is repeating after every $a$ occasions, then the UE transmits on $K - \left\lceil \frac{b-1}{a} \right\rceil a$ occasions, whereas in the proposed case, the UE transmits on $K - b + 1$ occasions, which is greater than the former numerical quantity.

It has to be noted that gNB can run into trouble if it detects some other interfering transmission before the late arrived first repetition or if the first repetition is not detected. In such erroneous scenarios, the gNB's understanding of the first repetition may change, and it will try to decode the wrong transmission or non-first repetition as the first repetition. Due to this, gNB may not be able to decode the first repetitions and probably the rest of the repetitions due to wrong RV understanding from gNB's point of view. However, a smart gNB can do blind decoding, and if it detects any repetition with some RV, then the rest of the repetition will follow the pattern and can be decoded readily without any blind attempts.

### 3.3 Common NACK feedback

An uplink transmission may fail depending on the propagating medium, transmission parameters, and hardware properties. It is plausible that if the UE transmits an uplink transmission, it may not be detected at all or is misdetected by the gNB. Another type of erroneous transmission could be a case where the gNB detects the transmission energy but cannot decode the UE identification (UE ID); for instance, the Demodulation Reference Signal (DMRS) sequence of the transmission can be used as UE ID. In these types of cases, a common NACK feedback along with CG can help to resolute the errors with minimal latency.

If the gNB detects transmission energy but is unable to decode the transmission (e.g., the UE ID), gNB broadcasts a group-common NACK in the cell indicating the time-frequency grid allocation of an unknown detected transmission. This is known as common NACK feedback from the gNB to its UEs. Once a UE receives and decodes the broadcast message and check if it had transmitted on the indicated resource, and if it did, it means their transmissions are not successfully received by the gNB. The UE must proceed with its retransmissions. In the case of URLLC allocation, the common NACK feedback can be utilized with CG allocation, where an individual UE retransmits on its immediate CG period and bypass the SR-based grant which requires considerable latency.

To enable common NACK feedback, the receiver employs an energy detector which is a part of the DMRS (e.g., UE ID) detector. However, to estimate the impact of common NACK feedback, the energy detection, and DMRS detection can be considered as separate operational events with different associated probabilities where the latter is dependent on the former. Let us denote $P_T$ and $P_E$, the probability of transmission and the probability of detection of the transmission. Now for every detected transmission, the UE ID can be identified successfully with probability $P_D$. Some transmissions may be misdetected with the wrong UE ID and have associated probability $P_{MD}$. The rest of the detected transmissions may not bear any positive UE ID from the gNB's active UEs, and this probability can be derived as $P_T P_E (1 - P_D - P_{MD})$. Upon detecting the identifying transmission, the gNB broadcasts the common NACK feedback, and the probable UE attempts to retransmit in the nearest CG period. The probability with which the retransmission can be detected is $P_T P_E (1 - P_D - P_{MD}) P_{CN} P_E P_D$ where $P_{CN}$ is the probability of common NACK feedback by the UE. Given the extremely reliable targets of an order of five or six nines, the retransmission success for the cases with no detection, misdetection, and unknown detection becomes extremely important.

The unknown detections can be attempted to resolute with the employment of common NACK feedback.

### 3.4 Assistive shared resource

To provide a URLLC service to a UE, gNB can allocate a dedicated CG resource. However, due to sporadic, or semi-deterministic traffic arrival, the UE may not be able to transmit on all the occasions within a CG period. This means the achievable reliability would be less than it would achieve with transmissions in all the occasions, namely $K$ repetition occasions in a CG period. To overcome the issue of a fluctuating arrival, the period must be equipped $K_+$ extra occasions in addition to $K$ occasions, so that a UE can transmit reasonably well and can achieve the same reliability than it would have achieved with deterministic arrival with $K$ repetition transmissions. However, this extra dedicated repetition allocation $K_+$ can be a costly means to compensate for deficient reliability due to the late arrival in the period. This is discussed in [12], where for a typical Shannon rate-based utility, exhibiting concave behavior, the additional benefits reduce with the additional consumption of resources. It means, if a UE transmits with two repetitions, the reliability gain would be less than the double of reliability gain from a single repetition.

To avoid overallocation in a CG period, the network may look for cheaper sources to provide additional reliability for the late arrivals. One solution is that the UE utilizes an assistive resource shared among the group of UEs., e.g., resource equivalent to $K_+$ occasions are allocated per slot which is shared among $N$ UEs. In this resource, a UE can transmit the remaining repetitions in a grant-free manner which it could not transmit in the dedicated CG period, see Fig. 5. If multiple UEs try to access the same occasion, the collisions may happen, and the repetition(s) may render useless. The allocated shared resource can be derived in such a manner that the probability of success in the dedicated plus shared resources subject to collisions provides the target reliability. The optimization equation can help us to derive, for instance, the number of occasions per slot. In [13], only shared resource is assumed for CG, and the author derived the allocation subject to the MAC reliability requirement. The same procedure can be employed using hybrid allocation, where some repetitions may occur in the dedicated CG and the rest in the shared CG. Also, with the addition of assistive resources, the system can serve more URLLC traffic, as the load is partly served by dedicated (NR) and shared (NR-U) spectrum resources.

Briefly discussing the resource $K_+$ derivation procedure, let us denote $P_a$ the probability of arrival in the repetition occasion in a dedicated CG which depends on the chosen traffic model. If the arrival misses all the occasions in the dedicated CG, then instead of transmitting all the repetitions in the shared CG, it postpones its transmission to the next CG, and this ensures at least one repetition occurs in the CG for the data transmission. The remaining repetitions will be done in the shared CG where each transmission has an associated collision probability. Thus, an error function can be derived as $P_{e-D}(P_a, K, \gamma) P_{e-S}(P_a, K, \gamma, K_+, N)$ where $P_{e-X}(\cdot)$ denotes the error probability for an allocation $X$, where $X = D$ is dedicated CG allocation, and $X = S$ is shared CG allocation and $\gamma$ the SINR of the transmission in the selected occasion. In the shared CG, the error probability is a function of UE population $N$, where the error probability in the shared CG increases with an increase in the UEs' population [13]. The value $K_+$ can be computed that deliver an error target of the order of $0.00001$.

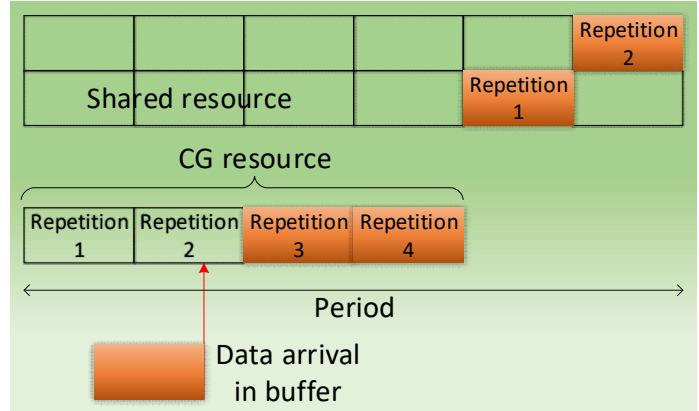

**Figure 5:** A UE is configured with $K = 4$ repetitions. To guarantee the reliability, if the UE is unable to transmit $K$ repetitions due to late data-arrival, it uses a shared resource.

### 3.5 Complementary TDD pattern

The CG can be allocated in TDD or Frequency-division duplexing (FDD) spectrum. However, with TDD, if certain slots/symbols are configured in the downlink direction, then part of TOs belonging to $K$ TOs downlink slots/symbols in a TDD pattern may deem invalid. That may result in a UE being forced to transmit less than the required $K$ repetitions. This can happen if there is a restriction with the TDD pattern and a single carrier is used.

To prevent this, an additional spectrum or carrier can be configured which acts in complement to the conflicting TOs in the primary spectrum, preferably on the NR-U spectrum. In the additional spectrum, uplink TOs can be allocated, when the UE has invalid TOs in the primary spectrum. Normally, the gNB may have a uniform TDD pattern across the spectrum in a geographical area, and it is cumbersome for the gNB to bifurcate the spectrum into different TDD patterns. One way to access the additional spectrum is the NR-U spectrum which is free and in abundance [11]. The only constraint is to observe channel sensing norms. Hence, if a UE transmits $X$ repetitions and is unable to transmit $K - X$ repetitions due to the conflicting TDD pattern in the NR spectrum, then in those conflicting slots, UE can be configured with repetitions in the NR-U spectrum. A similar method is specified in [14] where the FDD spectrum is flexibly allocated among the cellular and D2D UEs for providing URLLC services.

## 4 CONCLUSION

We presented an overview of the 3rd Generation Partnership Project standardization effort for New Radio configured grant specifications in Release 15 and 16 to support ultra-reliable and low latency services. Configured grant allocation allows a user to transmit its uplink data on pre-configured resources without the need for a scheduling request. Coupled with shorter symbol size due to new numerology in New Radio, a configured grant can provide ultra-low latency access. New Radio has specified two versions of configured grant allocations, one is for licensed spectrum where it aims to support extreme reliability, and the other is for unlicensed spectrum, which focuses to combat channel access failure due to competition. Next, in the remainder of the article, we discussed the proposed enhancements for future releases targeting traffic variance and spectrum efficiency, e.g., time-gap between repetitions, group-common feedback, assisted shared resource usage, flexible repetition transmission, compliment time-division duplexing patterns. These pluralities of enhancements can equip configured grant to serve semi-deterministic traffic for ultra-reliable services.

# 6 BIOGRAPHY

**Majid Gerami (majid.gerami@ericsson.com)** received his M.Sc. and Ph.D. degree from KTH (The Royal Institute of Technology), Stockholm, Sweden, in Electrical Engineering. Since December 2016, he is working with Ericsson Research, Lund, Sweden. His research interests include information theory, signal processing, channel coding, and network coding.

**Bikramjit Singh (bikramjit.b.singh@ericsson.com)** received his M.Sc. degree from Aalto University in Communication Engineering in 2014. Currently, he is with Ericsson, Jorvas, Finland, and pursuing doctoral studies at Aalto University, School of Electrical Engineering, Finland. His research interests include futuristic 6G resource allocation, 5G URLLC services, TSN systems, contention-based access, and game-theoretic spectrum sharing. He is an author of more than a dozen articles and conference papers and has filed over 100 patents in the areas of wireless communications.